\documentclass[12pt]{iopart}

\usepackage{epsfig}

\begin{document}

\title[GHz QKD]{GHz QKD at telecom wavelengths using up-conversion detectors}

\author{R.~T.~Thew$^{1}$\footnote[1]{To whom correspondence should be addressed.}, S.~Tanzilli$^{1}$, L.~Krainer$^{2}$, S.~C.~Zeller$^{2}$, A.~Rochas$^{3}$, I.~Rech$^{4}$, S.~Cova$^{4,5}$, H.~Zbinden$^{1}$ and N.~Gisin$^{1}$}

\address{$^{1}$ Group of Applied Physics, University of Geneva, 1211 Geneva 4, Switzerland}

\address{$^{2}$ Ultrafast Laser Physics Group, Institute of Quantumelectronics, ETH Zurich, Switzerland}

\address{$^{3}$ idQuantique SA, Chemin de la Marbrerie 3, 1227, Geneva, Switzerland}

\address{$^{4}$ Dipartimento Elettronica e Informazione, Politecnico di Milano, 20133 Milano, Italy }

\address{$^{5}$ Micro Photon Devices, via Stradivari 4, 39100 Bolzano, Italy}

\begin{abstract}
We have developed a hybrid single photon detection scheme for telecom wavelengths based on nonlinear sum-frequency generation and silicon single-photon avalanche diodes (SPADs). The SPAD devices employed have been designed to have very narrow temporal response, i.e. low jitter, which we can exploit for increasing the allowable bit rate for quantum key distribution. The wavelength conversion is obtained using periodically poled Lithium niobate waveguides (W/Gs). The inherently high efficiency of these W/Gs allows us to use a continuous wave laser to seed the nonlinear conversion so as to have a continuous detection scheme. We also present a 1.27\,GHz  qubit repetition rate, one-way phase encoding, quantum key distribution experiment operating at telecom wavelengths that takes advantage of this detection scheme.  The proof of principle experiment shows a system capable of MHz raw count rates with a QBER less than 2\,\% and estimated secure key rates greater than 100\,kbit/s over 25\,km.

\end{abstract}

\pacs{03.67.Dd, 03.67.Hk, 42.65.Ky, 42.65.Wi}



\maketitle

\section{Introduction}
The speed with which quantum key distribution (QKD) has progressed in recent years has been remarkable. From the original idea in 1984 \cite{Bennett84a}, to the first key exchange in 1992 \cite{Bennett92d} and then outside the lab by 1995 \cite{Townsend93a, Muller95a}, to commercial fruition  in 2001 \cite{RetailQKD04a}, only 20 years have passed. Unfortunately the speed of the actual key exchange  remains quite slow. We have seen some advances towards faster QKD with clock synchronisation over 1\,GHz  already achieved \cite{Bienfang04a, Gordon05a}. In both of these cases attainable distance was limited by the choice of wavelength and the rate of the quantum channel was limited by  a relatively wide detector response (i.e. fairly large timing jitter). Further progress has also led to the realisation that single or two photon sources are not required for security, and that weak pulses suffice. There have also been other more fundamental  attempts to find different protocols that allow for an increase in the quantum communication rates for weak pulse schemes like decoy state QKD \cite{Hwang03a, Wang05a, Lo05a}, the SARG protocol \cite{Scarani04a}, and the protocol recently proposed by Stucki {\it et al.} \cite{Stucki05a}.

Our goal is to increase the QKD rate for telecom wavelengths. To this end, we exploit planar Silicon (Si) SPAD devices \cite{Cova89a,Rochas03a} with very low timing jitter, of the order of 40\,ps, that are capable of MHz regime photon counting and have low noise and afterpulse probabilities. This first of all requires that we convert from telecom transmission wavelengths to the Si detection band. To achieve this we use nonlinear sum-frequency generation (SFG), or up-conversion in a periodically poled Lithium niobate (PPLN) waveguide (W/G). Takesue {\it et al.}  \cite{Takesue05a}  have recently incorporated a similar detection scheme into differential phase shift QKD, increasing both rate and distance for secure QKD, though using standard Si SPAD with larger timing jitter, of the order of 400\,ps and more \cite{Bienfang04a,Gordon05a}. By incorporating the low jitter SPADs in the experimental apparatus we can further increase the secure bit-rates and obtain greater levels of stability for interferometric based schemes.  What we show in this article is that whilst there are important limitations for the optics, these do not restrict us from making significant increases in the quantum bit rate for long distance (telecom wavelength) QKD with a careful choice of system parameters and components. 

This article is organised in the following manner. We first introduce the basic scheme used to perform weak pulse phase encoding QKD. We then briefly describe the two different Si SPADs that we have used before we discuss the results of a wavelength conversion process for single photon detection of telecom wavelength photons. The combined characteristics of these Si-SFG detectors are then highlighted in a proof of principle experiment where they are utilised in a 1.27\,GHz one-way QKD set-up.


\section{Fast phase encoding QKD}

\begin{figure}[h!]
\begin{center}
\epsfig{figure=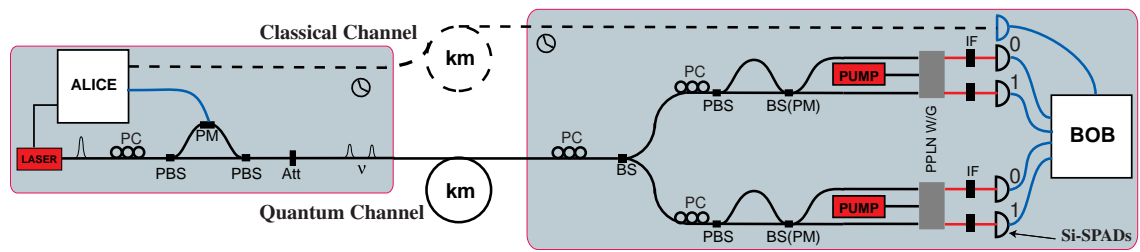,width=150mm}
\caption{The proposed scheme for GHz, weak pulse, phase encoding, QKD. }
\label{fig:RQKDSchematic}
\end{center}
\end{figure}

We wish to implement weak pulse phase encoding QKD for both BB84  \cite{Bennett84a} and SARG \cite{Scarani04a} implementations where both have quantum and classical channel repetition rates of 1.27\,GHz. We first present a general view of the experimental set-up, as shown in figure \ref{fig:RQKDSchematic}, to both highlight some of the crucial elements and at the same time to remind the reader of the protocols and how we have implemented them here. A weak coherent pulse from a laser is passed through an unbalanced Mach-Zender interferometer where a phase modulator (PM) [Covega 10\,GHz] acts on one arm. This introduces one of four relative phases (these are the two phases for each of the two bases) between the two outgoing probability amplitudes associated with the short and long paths, that encode the qubit. In these scenarios we consider the qubits to be encoded such that a logical "0" corresponds to the short path or first "time-bin" and the logical "1" by the long path or second "time-bin". The qubits are then sent over some distance from Alice to Bob. On Bob's side we have a choice of two approaches: either we use one interferometer with another phase modulator to choose the measurement basis with two detectors; or, we use two interferometers, where each interferometer defines one measurement basis, randomly chosen via a beam splitter (BS), and we use four detectors, as depicted in figure \ref{fig:RQKDSchematic}. In either case we can attribute a result dependent on which detector fired. Alice and Bob then communicate classically so as to reconcile their choices and  thus share a raw key.  

Let us be more precise and outline and compare the BB84 and SARG protocols. BB84 is perhaps better known and the protocol works as follows:- Alice prepares one of four states that belong to one of two conjugate bases. Bob measures in one these predefined bases and announces the basis. They then perform reconciliation which simply means that if Bob measured in the same basis as Alice prepared her state then they keep the result as their bit. In the case of SARG one can use exactly the same hardware as only the reconciliation process differs. Alice prepares the same states, however, conversely to BB84, Bob announces the result of his measurement  and not the basis. The two protocols are completely symmetric and for the later it has the immediately obvious security benefit (against photon number splitting attacks) that even if Eve had stored a photon in her quantum memory, she still doesn't know which basis to measure in! Importantly, they only agree to the bit if Bob measures in the opposite basis to the one Alice used to prepare the state, and he only obtains the correct result half the time. The increased security allows one to use a larger mean photon number, though as we see, the price one pays is that we only keep one result in four after sifting as opposed to half for BB84  \cite{Scarani04a}.

If we now look at the detection system, we note that the W/Gs that are used for the SFG are polarisation sensitive, i.e. the process requires that the input photons have a specific polarisation with respect to the optic axis of the detection set-up, as we will see. This could be seen as a down side, but if  we need to control, in part, the polarisation of the input signal, we might as well do it before the interferometer as after it! By using polarising beam splitters (PBSs) at the exit and the entrance of Alice and  Bob's respective interferometers, we ensure that all signal photons arrive in coincidence with the clock signal in the same way that is done for the Plug\&Play scheme \cite{Gisin02a} and at the same time optimising the detector's efficiencies. Another subtle technical point that should be mentioned here when comparing these two approaches is that we do not need to consider extra loss for Bob if we add a phase modulator. One can simply balance the probability amplitudes, as we have done here, with the PC and PBS at the entry to Alice's interferometer such that the combined probability amplitudes for both interferometers are equal, i.e. the loss due to the PM is compensated.

There are several elements to this experiment that are advantageous for performing rapid QKD. The laser is mode-locked to provide not only a high repetition rate and photon extinction levels, but also the appropriate width, Fourier transform limited, pulses. The optical pulse width is perhaps the most crucial parameter limiting the repetition rate. The interferometers are made from polarisation maintaining fibre and have optical path length differences of 300\,ps. This extremely short time difference means that the interferometers are inherently more stable than most previous implementations (usually several ns) \cite{Townsend93a, Gobby04a}. This has been made possible due to the use of this detection scheme based on nonlinear W/Gs  and silicon (Si) SPADs with very low temporal jitter. We will now discuss in more detail each of these elements and how they contribute, starting with the detection system.


\section{Si-SFG single photon detectors}

 The idea of using SFG to facilitate a measurement in a bandwidth with better detection characteristics is not new \cite{Boyd68a}. However, it is only  with recent technological advances that this approach is being revisited to study the single photon detection regime for telecom wavelengths, as well as for longer wavelengths such as 4$\mu$m  \cite{Temporao05a}. In the telecom band, there are now several groups investigating this approach using commercially available single photon counting modules (SPCM-AQR), based on Si SPADs  with both bulk and waveguiding PPLN  crystals and either  continuous or  pulsed pump sources \cite{VanDevender04a, Roussev04a, Albota04a}. Whilst initial results have been promising, with respect to overall efficiencies, there have been significant problems with noise which we have not escaped either and will return to momentarily. More recently we have also seen experiments that have used this process but focused on the coherence of this conversion in the context of a quantum interface \cite{Tanzilli05a}. All of theses schemes so far  have incorporated standard Si SPADs with photon-timing jitter of the around 300-500\,ps. We have studied a combination of PPLN W/Gs [HC Photonics] and two different types of Si SPADs with low timing jitter.  The overall goal is to develop a compact and practical single photon detection system for telecom wavelengths with low noise, high temporal resolution and  a quantum efficiency comparable to current InGaAs SPADs. 
\begin{figure}[h!]
\begin{center}
\epsfig{figure=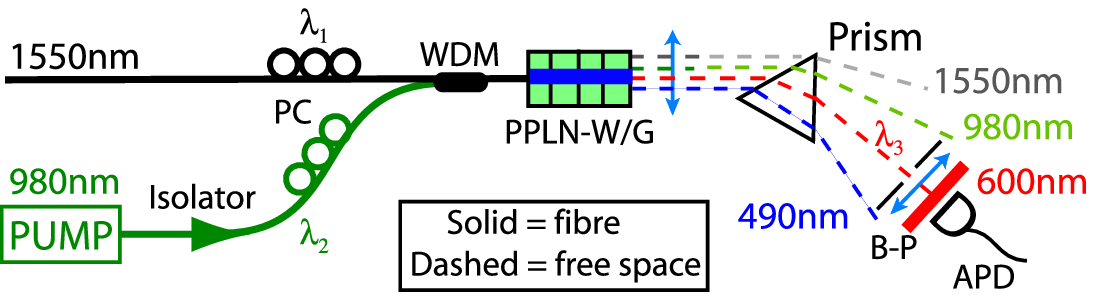,width=110mm}
\epsfig{figure=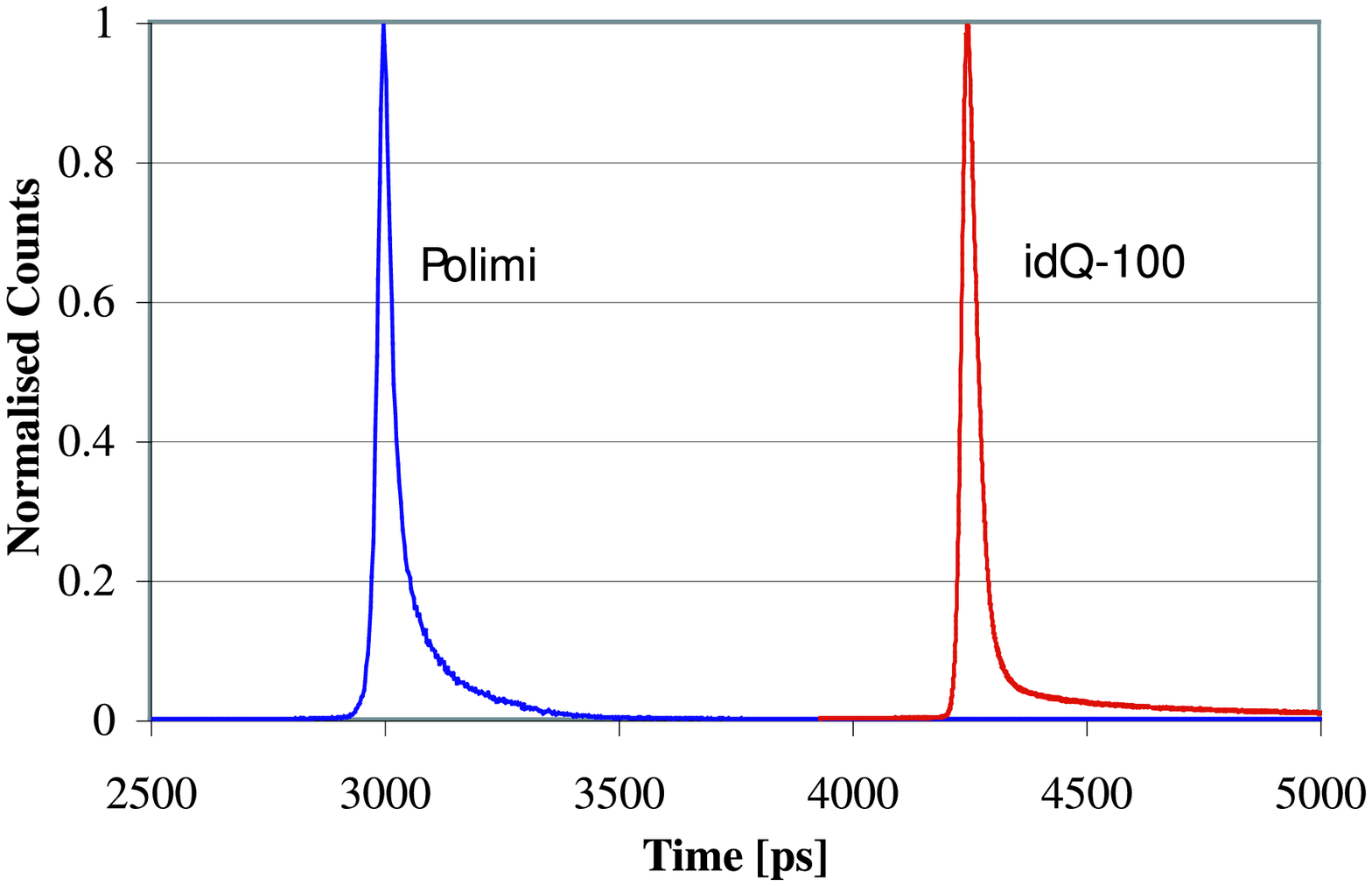,width=45mm}
\caption{(Left) Schematic of the Si-SFG hybrid detection scheme using PPLN waveguides and Si SPADs. (Right) Timing jitter for the two SPAD used. See text for details.}
\label{fig:DSchematic}
\end{center}
\end{figure}

Usually when working with SFG one operates in a regime  where there are many photons. Here we go to an extreme where we are really interested in converting a single photon  from one wavelength to another using a pump at a third wavelength to aid, or seed, the process. The energy conversion  process that we are interested in is $ 1550\,{\rm nm} + 980\,{\rm nm}  \rightarrow 600\,{\rm nm}$.  Because of the high conversion efficiency of these W/G devices \cite{Boyd68a}, we can use standard, and relatively inexpensive, 980\,nm diode lasers [JDS Uniphase]. This allows the Si-SFG detectors to be operated in a continuous mode. Whilst we don't foresee any problems for SFG at the single photon level, due to the high pump powers we are using here, there are other processes that can introduce noise to the overall process as we will discuss momentarily. Figure \ref{fig:DSchematic} shows the experimental arrangement for the detectors. A single photon at 1550\,nm is mixed with a large flow of 980\,nm pump photons at  a wavelength division multiplexer (WDM). The polarisation controllers (PC) ensure that all photons enter the PPLN W/G with the correct polarisation. The WDM  is pigtailed to the  W/G input. The  free space output  is then  collimated and passed through a dispersion prism to spatially filter the unwanted spectral components. After the W/G there is still a large flow of 980\,nm photons as well as 490\,nm photons that are generated by second harmonic generation (SHG) from the pump. It is crucial to filter these as the SPADs are relatively efficient at these wavelengths. The signal at 600\,nm is then focused through a bandpass filter (600-40\,nm) and onto the SPADs.

SPAD devices for accurate single-photon timing have been under development for many years at Politecnico di Milano (Polimi) resulting in various generations of silicon planar  CMOS (Complementary Metal Oxide Semiconductor) compatible technology \cite{Ghioni88a}. Their capability of photon timing with jitter down to 20\,ps has been demonstrated \cite{Cova89a} and the concurrent progress in electronics for the detector operation (integrated Active Quenching Circuit iAQC) have made it possible to develop compact single photon timing modules \cite{Rech04a}. More recently, Rochas {\it et al.} were able to design and fabricate SPAD devices in an industrially available silicon high-voltage CMOS circuit technology \cite{Rochas03a}. Both these SPADs [idQuantique id100  and Micro Photon Devices PDM20T] have features suitable to our purpose and have been employed in the experimental tests. They both exhibit timing jitter of 40\,ps, dark counts of less than 200\,Hz  with peak  detection efficiencies in the blue/green spectral range with an active area with 20\,$\mu$m diameter and afterpulse probabilities of 1\,\% at room temperature. The timing jitter for both of these devices is shown on the right of  figure \ref{fig:DSchematic}.

The SFG conversion efficiency can be shown to be $ \eta = N_{3}/N_{1}  =  \sin^2[(\eta_{norm} P_{2})^{1/2}L]$  \cite{Roussev04a}. Here $N_{1}$ is the number of 1550\,nm input signal photons and $N_{3}$ is the 600\,nm output signal photons for detection.  $P_{2}$ is the input power at 980\,nm  and $\eta_{norm}$ is a normalised internal conversion efficiency of the W/G. This is an idealised solution, which can reach 100\%, though in reality, for the overall efficiency, we need to consider coupling losses and internal transmission losses such that the conversion efficiency will be limited, in our case to around 30-35\%. Finally we need to include filtering and the efficiency of the Si SPAD itself, which will further reduce the overall detection efficiency.  

\begin{figure}[h!]
\begin{center}
\epsfig{figure=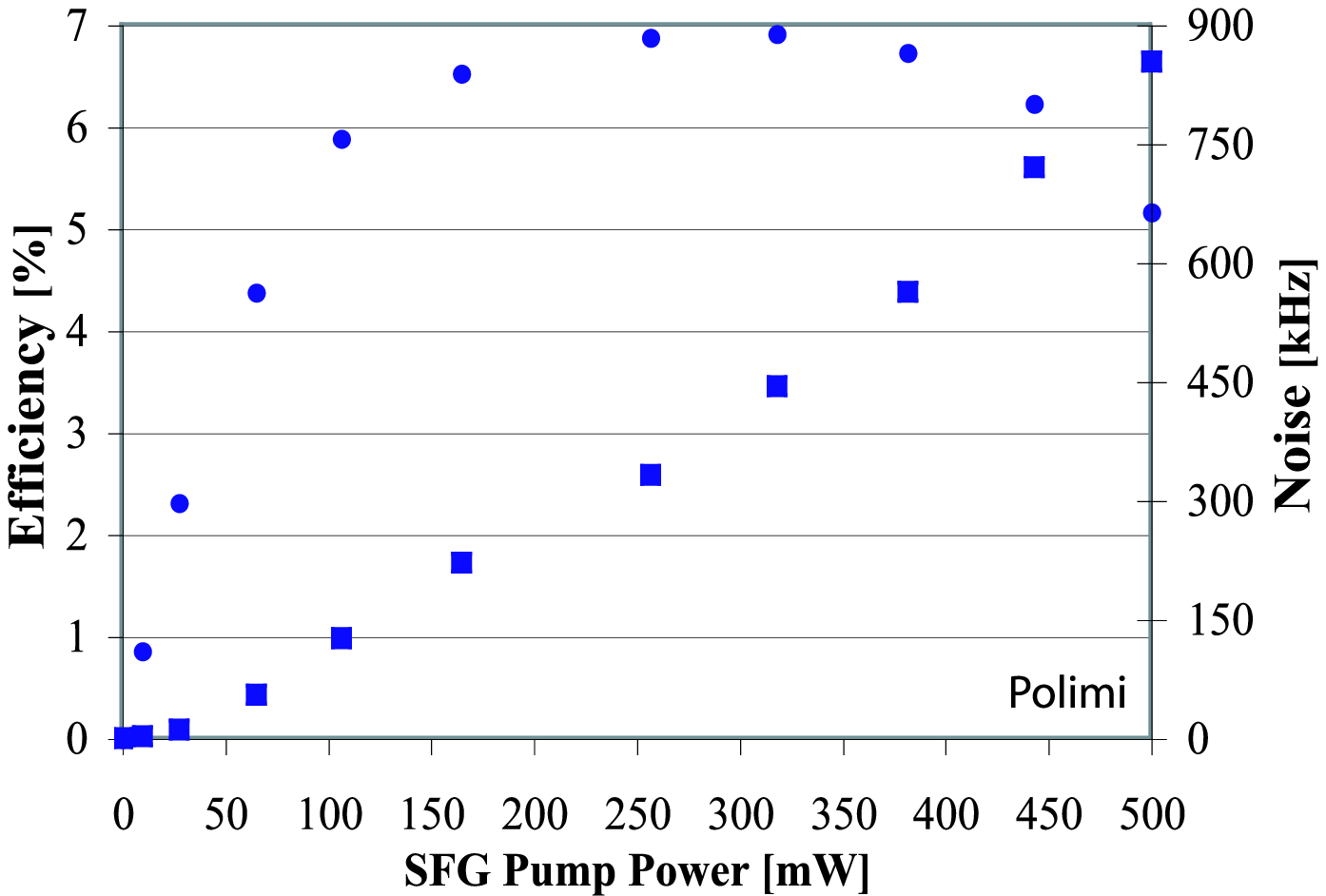,width=75mm}
\epsfig{figure=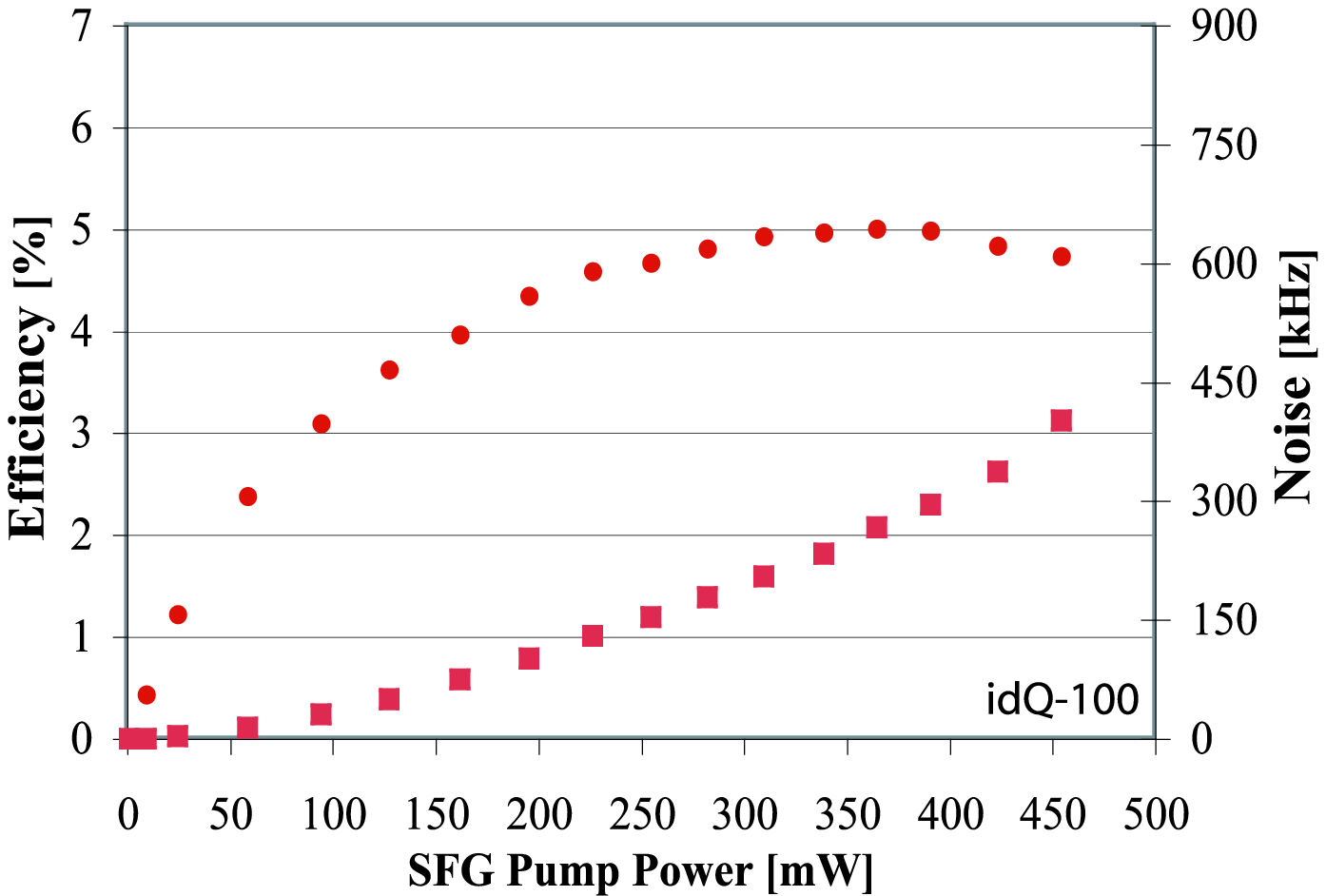,width=75mm}
\caption{The overall detection efficiency (circles) and noise response (squares) as a function of the input pump power. (Left) Polimi Si-SPAD. (Right) idQ Si-SPAD.}
\label{fig:VsPower}
\end{center}
\end{figure}
To characterise the detectors we first look at this conversion efficiency and noise as a function of the input pump power $P_{2}$ at 980\,nm.  To do this we have used a set-up as depicted in figure \ref{fig:DSchematic} with  two calibrated variable attenuators to reduce the 1550\,nm signal down to the single photon counting regime \cite{Comment1}. The results for both W/G-SPAD combinations under test are shown in figure \ref{fig:VsPower}. We find peak detection efficiencies of around 5-7\% which is in good agreement with the theory given the conversion, filtering and detection characteristics for each set-up. We note that the efficiency reaches a peak and then starts to degrade as the pump power increases. This is the normal response of this type of process  where we have so many photons at 600\,nm that it becomes efficient to reconvert them  back to 1550\,nm, i.e. the inverse process of parametric down-conversion (PDC).

We also see the noise characteristics for these Si-SFG detectors is not only significant but nonlinear as a function of the pump power. The noise is measured concurrently with the efficiency by blocking the input signal and is simply the number of detection events one obtains without an input. Whilst one would expect the large number of pump photons could contribute, as the SPADs are still sensitive at these wavelengths, suitable filtering can remove this. The response for this should also be linear, but here there is clearly a nonlinear component that  is not normal. A further study as to the origins of this noise is currently underway as there are several possibilities. Fluorescence and Raman scattering have already been suggested in similar experiments \cite{VanDevender04a, Roussev04a, Albota04a}. We would also like to put fourth another possibility for the case of PPLN crystals and specifically W/Gs. 

Phase matching for different processes in PPLN W/Gs is achieved by changing the size of the poling periods \cite{Sutherland96a}. As mentioned previously, SFG is the inverse of PDC, and we have also observed SHG in these W/Gs, so we can start to imagine a large family of possible conversion processes. One allowable process is for PDC, where the 980\,nm pump generates photon pairs at 1550\,nm and 2665\,nm. This 1550\,nm photon can then in turn be converted to 600\,nm which would produce a quadratic noise response, as we see. Unfortunately we cannot tune to another wavelength as the 2665\,nm photons are not guided, but emitted into the W/G substrate. In doing this they can satisfy Cerenkov phase matching conditions which results in a broad band generation of photons around 1550\,nm \cite{DeMicheli97a} which we could observe experimentally.

In our case we are interested in QKD and as such the efficiency-noise relationship is fundamental. Even though we have relatively high levels of noise we find, with both W/G-SPAD combinations, that we can reduce the efficiency to around 2\% with less than 20\,kHz of noise. We can then take further advantage of the low jitter  of the SPADs, as illustrated on the right of figure \ref{fig:DSchematic}, by working with detection windows of only a few hundred ps. If we do this, we see that we have an effective noise probability  per gate which is comparable with current InGaAs SPADs, though we now have a detector that can operate in the MHz counting regime.


\section{The chromatic dispersion limit - Fourier transform limited source}

 We have overcome several problems on the detection side with the Si-SFG detectors, so now let us look at the optical pulses. We need to have relatively short pulses to take advantage of the temporal resolution the SPADs provide. However, we will be using fiber optic transmission and as such chromatic dispersion needs to be carefully considered \cite{Fasel04a}. Chromatic dispersion in standard optical fibers at 1550\,nm is around 17\,ps\,nm$^{-1}$\,km$^{-1}$. There is a spectral bandwidth between 80\,-\,200\,pm, that provides transform limited pulses of  40-100\,ps, that is ideal for transmission up to 50\,km in fibre. Otherwise, one either starts with pulses that are too short and quickly have extremely large pulse widths due to chromatic dispersion, or we simply start off with pulses that are too large to take advantage of our small jitter SPADs. This is a difficult regime in which to find telecom wavelength lasers that are both, in the GHz repetition range, and have short, but not too short, pulses.

In this instance we have chosen to work with a custom made mode-locked telecom wavelength (1550) laser operating at a repetition rate of 1.27\,GHz (Ulltrafast laser physics group, ETH Zurich). There are many compelling reasons to use a pulsed laser directly as an optical source in telecommunications systems, both classical and quantum. First, this eliminates the need for a modulator to create the pulses and thereby can simplify system architecture.  Secondly, the extinction ratio of pulsed lasers is typically very good and much higher than for modulated cw sources improving the system signal-to-noise ratio. Furthermore, the pulses produced are transform limited and hence present a much better proposition for fibre optic transmission. Passively mode-locked, diode pumped solid-state lasers have been presented with repetition rates up to 50\,GHz around 1550\,nm  \cite{Zeller04a}, and up to 80 GHz in the 1\,$\mu$m wavelength region \cite{Lecomte05a}. With novel pump diodes a previously demonstrated 160\,GHz laser could also easily be extended to diode pumping \cite{Krainer02a}. 
\begin{figure}[!h]
\begin{center}
\epsfig{figure=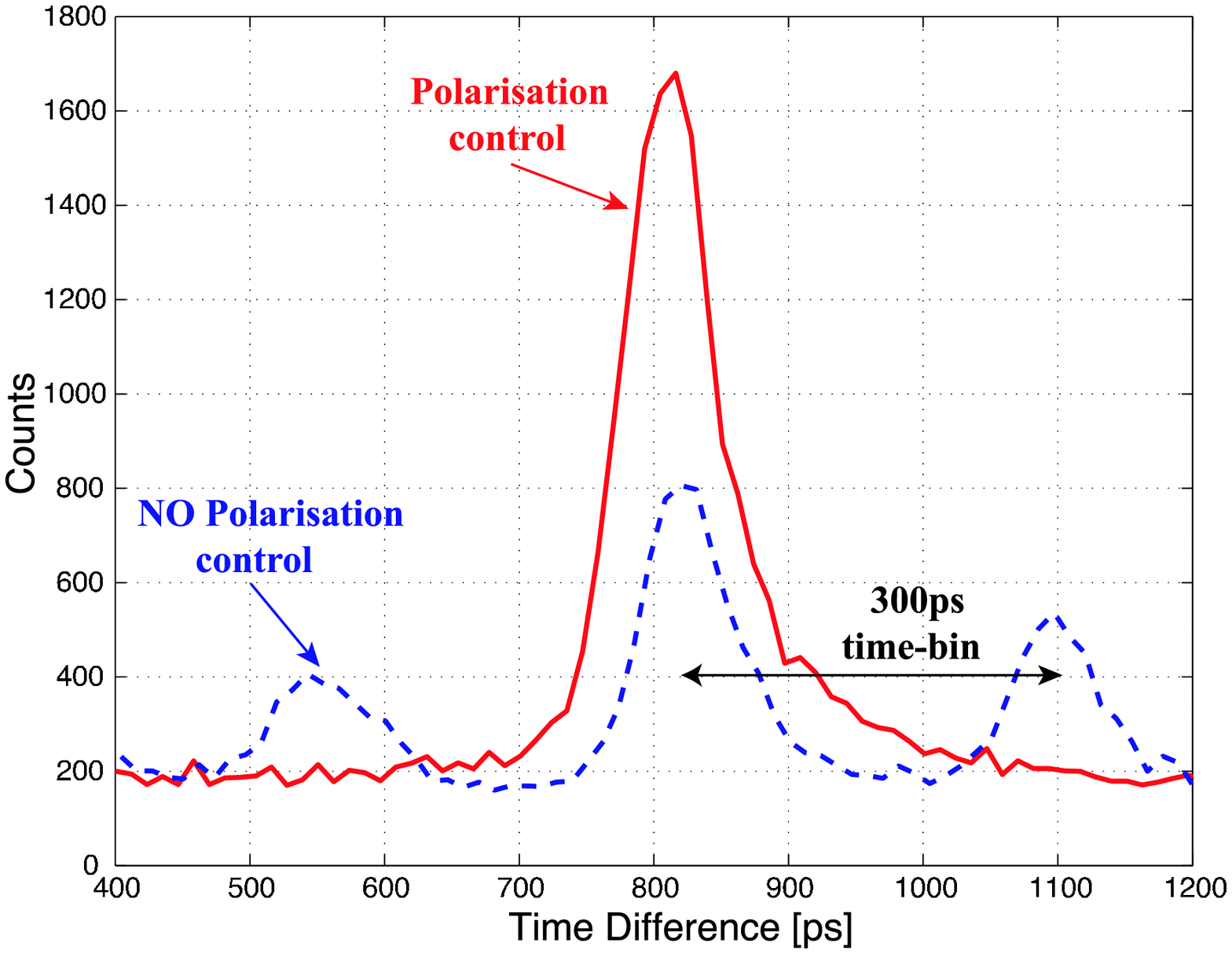,width=75mm}
\epsfig{figure=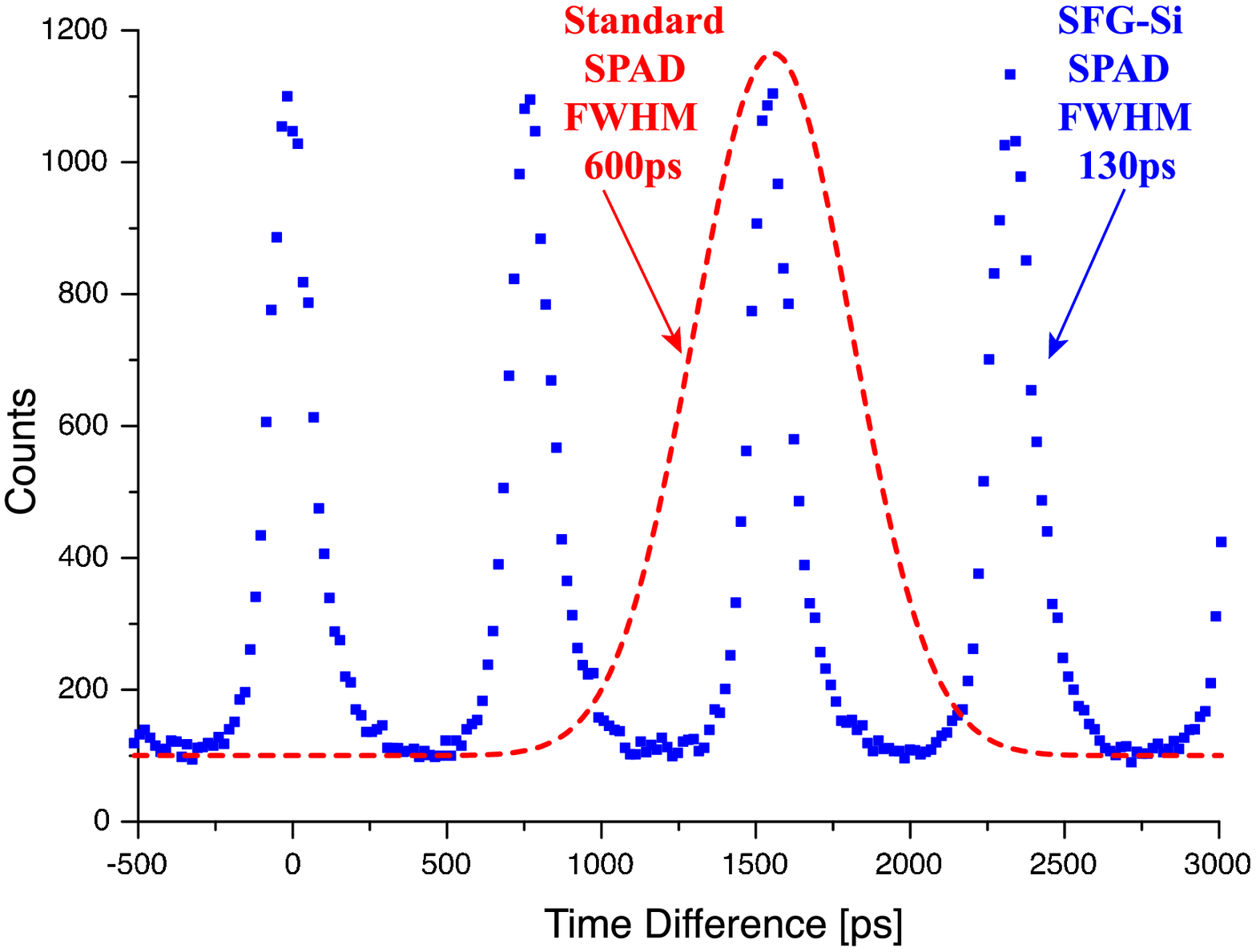,width=75mm}
\caption{(Left) Time difference histogram showing the effects of controlling the polarisation to remove the satellite peaks. (Right) Extended view of a train of pulses highlighting the side peak extinction and the superior temporal resolution one would expect with respect to standard Si-SPADS.}
\label{fig:Train}
\end{center}
\end{figure}

The laser used in this experiment has an Er:Yb:glass gain medium  that is pumped by telecom-grade 980\,nm laser diode, which typically delivers 300\,mW of pump power in a diffraction-limited beam. Most of the pump radiation is absorbed by the ytterbium dopant and then transferred to erbium ions, generating gain in the 1550\,nm region. Passive mode-locking is achieved by placing a semiconductor saturable absorber mirror (SESAM) at one end of the cavity \cite{Keller92a,Keller96a,Keller03a}.   An intracavity etalon is included to adjust the laser wavelength and to set the pulse duration to the desired value. Typical pulse durations are in the range of 1\,ps up to nearly 20\,ps, while the average output power is around 10\,mW. Further external spectral filtering [AOS fibre Bragg grating - 100\,pm] is used which corresponds to Fourier transform limited pulses of around 80\,ps.

\section{Proof of Principle for rapid QKD}

Let us now look at the first proof of principle results. The time-of-arrival histogram on the left  of figure \ref{fig:Train} illustrates the type of signal with which we are dealing. This figure shows the two cases, where the polarisation is controlled, or not, before entering Bob's interferometer. We see that when the polarisation is controlled the satellite peaks (non-interfering short-short and long-long events for the interferometers) can be made to disappear. We control the polarisation between the two interferometers with a polarisation controller (PC), also illustrated in figure \ref{fig:RQKDSchematic}. This approach reduces the noise level due to pulses spreading into adjacent time-bins.  If we do not control Bob's input polarisation and the channel is completely depolarised, the count rate is reduced to its normal level, i.e. we have all three peaks and reduce the bit rate by a half.  

We note here that the histogram is generated with start and stop signals given by the two SPADs and not with one SPAD and the clock. This is due to the clock being too fast for the available time-to-amplitude converter [Tenelec TC863]. As such the FWHM of these peaks includes the contributions from both SPADs, the optical pulse, and all of the associated electronics.  It is important to note that the following measurement data, for the QKD, is generated independently to this and corresponds to coincidence events between the clock and each of the SPADs. These widths govern both, how quickly we can send successive pulses and how large the path length difference can be in the interferometers. Even in an extended view of the pulse train that we are using, as shown on the right of figure \ref{fig:Train}, we see that the pulses are clearly defined and that the side peaks are suppressed. Furthermore, at 25 and 50\,km we only observe a broadening of these histogram peaks of around 40 to 80\,ps respectively, as expected. The dashed curve with a FWHM shows the type of response we would expect from a standard Si-SPADs.


 In this proof-of-principle experiment we have not sent random bit trains but simply the same bit many times. Measurements were made with two different lengths of fibre, 25\,km and 50\,km. In this first instance our goal is to highlight the advantages of the timing jitter and count rates afforded by the SFG-Si SPADs. Therefore, for each run, we set the phase, take an average, then set the next phase. The raw data for a series of these measurements is shown in \tref{Tab:RawResults}. The results are shown for each detector (D$_1$, D$_2$) and are averages over a one minute interval, for two different states which are defined by Alice's choice of phase. The first value in each case, either result "0" or result "1", being the correct or True (T) value, followed by the wrong or False (F) value. We clearly see the advantages of using the SFG-Si detection scheme as it is now possible to have count rates R of over 2\,MHz. 
\begin{table}[h!]
\begin{center}
 \caption{Raw data for the coincidences (SPAD + Clock) for the 2 protocols and two distances (25 and 50\,km). See text for details.}\vspace{2mm}
 \begin{tabular}[!h]{|c|c|c|c|c|c|}\hline\hline
  &True / False & BB84-25  & SARG-25 & BB84-50 & SARG-50 \\ \hline  \hline
"0" & D$_1$(T) / D$_2$(F)  & 763\,k / 11.6\,k  &  2.24\,M / 17.9\,k  & 117\,k / 9.54\,k &  636\,k / 12.1\,k  \\ \hline
"1" & D$_2$(T) / D$_1$(F)   & 660\,k / 14.5\,k  &  1.84\,M / 19.3\,k  & 104\,k / 11.5\,k  & 558\,k / 16.9\,k  \\ \hline
 \end{tabular}\label{Tab:RawResults} 
\end{center}
\end{table}

We now need to think about the QBER. The QBER is simply the ratio of false events to the total number of events after sifting (reconciliation) and for the two cases we find,
\begin{eqnarray}\begin{array}{ccccc}
\hspace{-2cm}{\rm QBER(BB84)} & =  &\frac{(1/2)(P_{opt} P_{phot} + P_{dark})}{(1/2)(P_{phot} + 2P_{dark})}\hspace{0.6cm} \approx P_{opt} + \frac{P_{dark}}{P_{phot}} 
& \equiv &Q_{opt} + Q_{det}\label{eq:qber1}, \vspace{2mm}\\ \vspace{1mm}
\hspace{-2cm}{\rm QBER(SARG)} & = & \frac{(1/2)(P_{opt} P_{phot} + P_{dark})}{(1/4)((1+ 2P_{opt}) P_{phot} + 4P_{dark})} \approx 2P_{opt} + 2\frac{P_{dark}}{P_{phot}} 
&\equiv &  2(Q_{opt} + Q_{det})\label{eq:qber2},
\end{array}
\end{eqnarray}
where $Q_{opt}=(1-V)/2$, ($V\sim 1$ being the visibility). We have also assumed that $P_{dark} << P_{phot} = \mu\,\eta t <<1$. As we initially said, there are two possibilities for Bob, one interferometer and two detectors or two interferometers and 4 detectors. The two cases do not give the same results, so it would be interesting to see how this varies for the two different protocols. In the case of four detectors we find that the component $Q_{det}$ increases by a factor two. It must be remembered that Q$_{det}$ is calculated using the optimal $\mu$ for each protocol, so things may not appear to be as big a problem for SARG as one might first think. The theory uses the following experimentally verified values, Q$_{opt} \approx 0.5$\,\%, a detection efficiency of 1.2\,\% and a detector dark count probability $P_{dark} \approx 7 {\rm x} 10^{-6}$ per gate. The gate width is defined by the electrical pulse widths of the clock and the Si-SPAD signals which are around 300\,ps. The experimental QBER is calculated simply as the average number of wrong counts divided by the average number of total counts for both detectors (using the results in \tref{Tab:RawResults}), with an extra factor of two in the case of SARG to account for the greater sifting losses. 

Once we have determined the errors we can then think about estimating the secure key rate $R_{sk}$ that we could expect. Firstly, to do this we need to be clear about the assumptions that we are using to analyse the security of the system. Obviously this is a weak pulse encoding scheme, and we assume we are working in a trusted device scenario where Eve doesn't have access to the detectors, i.e. she cannot change Bob's detector efficiencies or dark counts. In this instance we are not considering coherent attacks, and as such, Eve only performs individual attacks, or photon number splitting attacks when there are two or more photons per pulse. The reduction in the raw key rate for both cases  scales as,
\begin{eqnarray}
R(1-H(QBER) - I_{Eve})P_{sift}
\end{eqnarray}
where $H$ is the Shannon entropy, $I_{Eve}$ is Eve's information, and $P_{sift}$ = 1/2 for BB84  \cite{Niederberger05a} and $1/4(1 + Q_{opt} + 2Q_{det})$ for SARG \cite{Branciard05a}. Eve's information  reduces to $I_{Eve}\approx \mu/2t \approx 1/2$ for BB84 \cite{Niederberger05a}, and $I_{Eve} \approx I_1 + (1-I_1)\mu^2/12t \approx 0.6$ for SARG, where $I_1\approx 0.4$ \cite{Branciard05a}. In each case the optimal $\mu$ (mean number of photons per pulse) is used, BB84 where $\mu_{opt} = t$, and SARG where $\mu_{opt} = 2\sqrt{t}$ ($t \equiv$ transmission). For a detailed discussion concerning the security of these schemes, the reader is referred to \cite{Niederberger05a} for BB84 and  \cite{Branciard05a} for SARG.

\begin{table}[!h]
\begin{center}
\caption{Results for GHz QKD for two different fibre lengths using both BB84 and SARG protocols. R\,=\,raw key rate, S\,=\,secure key rate.}\vspace{2mm}
\begin{tabular}[!h]{|c|c|c|c|c|c|c|}\hline\hline
{\rm Protocol} & km & $\mu$   & R [Bit/s ] & QBER(exp) [\%] & QBER(th.) [\%]& S [Bit/s ]\\ \hline  \hline
BB84 & 25 & 0.286 & 710\,k  & 1.84 & 1.56 & 135k \\ \hline
SARG & 25& 1.064 & 2.04\,M  & 1.82 & 1.62 & 140\,k \\ \hline
BB84 & 50  &  0.101 & 110\,k  & 9.51 & 9.21 & 2k \\ \hline
SARG & 50 & 0.640 & 590\,k  & 4.71 & 4.11 & \,20k \\ \hline
 \end{tabular}\label{Tab:Results} 
\end{center}
\end{table}
In \tref{Tab:Results} we see the average results for the raw bit rates, the QBER and an estimate of the secret key rate, given R and the QBER, for the case of one interferometer and two detectors. The good correspondence between the QBER for theory and experiment implies that the error due to chromatic dispersion effects is minimal, indeed a Q$_{disp} \approx 0.3$\,\% for the 50\,km transmission would bring the two inline. Whilst with this set up we are capable of performing QKD using either protocol over these distances, there are definite advantages to using SARG. At 25\,km we have a raw count rate of over 2\,Mbit/s for SARG compared with 710\,kbit/s for BB84, both with similar QBER. At this distance there is not a great difference in terms of the QBER or final secret key rate. However, at 50\,km we have 590\,kbit/s and 110\,kBit/s respectively, and whilst the QBER for BB84 has increased significantly, and hence we extract only a relatively small amount, SARG still has a respectable QBER of less than 5\,\%, and we estimate a secure key rate of around 20\,kbit/s, an order of magnitude improvement.


\section{Conclusion}

We have presented two primary results, a SFG-Si telecom single photon detector that has low jitter and is capable of MHz count rates, and a GHz QKD system that is optimised to use this detection scheme.  The Si-SFG hybrid detector has a single photon detection efficiency of over 5\%, though with a significant noise problem. This requires further study to determine to what level this noise problem can be minimised. However, as it stands the temporal resolution and afterpulse effects are better than standard InGaAs APDs for telecom wavelengths and we can also use these detectors in a continuous, or free running, mode. The phase encoding QKD scheme has been shown to work over distances of 25\,km and 50\,km of standard telecom fibre. Importantly, these results show that the bit rate for long distance QKD can be substantially increased before there are any significant optical restrictions.  The system has shown that it is capable of MHz raw bit rates and estimated secure key rates greater than 100\,kbit/s over distances of 25\,km and 20\,kbit/s over 50\,km. Indeed, with a foreseeable reduction in detector noise, and minor increases in detection efficiency, high secure rates up to distances of 100\,km should be possible.

\ack

The Geneva group would like to thank V.~Scarani and C.~Branciard for helpful discussion concerning QKD theory and also C.~Barreiro and J-D. Gautier for their technical assistance. They also acknowledge financial support from the European IST programme RamboQ  and the Swiss NCCR "Quantum Photonics".  The Zurich group acknowledges the support of the Hasler Stiftung.


\begin{thebibliography}{<20>}
\bibitem{Comment1} Two attenuators are used so that both can be operated well within their calibrated linear regime.
\end{thebibliography}
\end{document}